\documentclass[3p, authoryear]{elsarticle}

\usepackage{latexsym}
\usepackage{graphicx}
\usepackage{epsfig}
\usepackage{amssymb}
\usepackage{amsmath}
\usepackage{amsfonts}
\usepackage{latexsym}
\usepackage{graphicx}
\usepackage{amsfonts}
\usepackage{dsfont}

\def\bd{{\bf d}}

\def\bx{{\bf x}}

\def\R{{\rm I} \! {\rm R}}

\newfont {\car} {eufm10}
\newfont {\Lar} {eufb10}
\newfont {\lar} {eufm5}
\newfont {\ab} {msbm5}
\newfont {\AB} {msbm10}
\newfont {\Lab} {msbm10}
\newtheorem{defin}{Definition}[section]

\newtheorem{rem}{Remark}[section]
\newtheorem{teo}{Theorem}[section]
\newtheorem{cor}{Corollary}[section]

\def\beq{\begin{equation}}
\def\eeq{\end{equation}}
\newcommand{\newint}[4] {\int_{#1}^{#2}\!\!#3\,{\rm d}#4}                   % correctly spaced integral
\def\bd0{{\bf 0}}

\newcommand{\pro}[2]{#1'(#2)}                                                                                                                                           % prime, one argument
\newcommand{\hist}[2]                           {#1^{t}(\bx,#2)}                                                                                    % history  of argument
\newcommand{\mint}[2]                                                                                                                                                                           % memory integral of kernel, vector
{#1_{0}#2(\x,t)+\newint{0}{\infty}{\pro{#1}{s}\hist{#2}{s}}{s}}

\begin{document}
\begin{frontmatter}

\title{On the asymptotic behavior of the quasi-static problem for a linear  viscoelastic fluid }
%\author{Mauro Fabrizio, Barbara Lazzari and Roberta Nibbi\\Dipartimento di Matematica - Universit\`a di Bologna}
 \tnotetext[t1] {Research performed under the
auspices of G.N.F.M. - I.N.d.A.M. and partially supported by Italian
M.I.U.R..}
\author[bologna]{Mauro Fabrizio}
\ead{fabrizio@dm.unibo.it}
\author[bologna]{Barbara Lazzari\corref{cor}}
\ead{lazzari@dm.unibo.it}
\author[bologna]{Roberta Nibbi}
\ead{nibbi@dm.unibo.it}
\cortext[cor]{Corresponding author}
\address[bologna] {Department of Mathematics, University of Bologna, 5 Piazza di Porta S. Donato,
40126 Bologna, Italy}
\begin{abstract}  
In this paper we study the quasi-static problem for a viscoelastic fluid by means of the concept of minimal state. This implies the use of a different free energy defined in a wider space of data.
The existence and uniqueness is proved in this new space and the asymptotic decay for the problem with non vanishing supplies is obtained  for a large class of memory kernels, including those presenting an exponential or polynomial decay.
\end{abstract}

\end{frontmatter}

\renewcommand{\theequation}{\thesection.\arabic{equation}}
\setcounter{equation}{0}
\section {Introduction}
When studying  materials with memory, the classical approach is based on the histories of the deformation gradient. In \cite{DD1}
it has been shown that different histories may lead to the same response of the material%
\begin{footnote}
{For completeness, we recall that a first contribution in this direction was presented in \cite{Banfi}.}
\end{footnote} and a new concept of state, relying on the minimal information required to determine the further behavior of the material, has been introduced for linear viscoelastic models.
Furthermore, it is well known that several free energies can be defined for materials with memory. The family %$\mathcal{F}$
of the free energies is a convex set which has a minimum and a maximum element $\psi_{min}$, $\psi_{max}$. It follows that, for any free energy $\psi$, the state domain $\mathcal{H}_{\psi}$, for  which $\psi$ is finite, is such that
$\mathcal{H}_{\psi_{max}} \subset \mathcal{H}_{\psi} \subset \mathcal{H}_{\psi_{min}}$.

In this paper, making use of the concept of minimal state, we study the quasi-static problem for a linear incompressible viscoelastic fluid and prove that it admits a unique solution belonging to $\mathcal{H}_{\psi_{min}}$ if the memory kernel satisfies only the restrictions imposed by the Laws of Thermodynamics and the data belong to the dual space of $\mathcal{H}_{\psi_{min}}$, which is the widest space that one can expect.

As for the long time behavior, many results have been established for the dynamic problem with memory kernels exhibiting exponential or polynomial decay, but in general for vanishing supplies.

Recently \cite{Messaoudi2008} has proposed a unified approach and proved that, in the evolutive problem with vanishing past histories and no external forces, the energy has the same type of temporal decay of the memory kernel, which is not necessarily an exponential or polynomial decay.

Here, we restrict ourselves to the quasi-static approximation of the problem and obtain a temporal decay similar to the one obtained by \cite{Messaoudi2008}, but in presence of supplies and in a wider space of initial past histories.

This is a first step in order to apply both the concept of minimal state and the unified approach
to more general problems in viscoelasticity.

The paper is organized as follows. In Section 2 we recall some properties of linear viscoelastic fluids and  introduce the concept of minimal state. In Section 3  we consider the quasi-static problem and establish its well posedness, while in Section 4  we present our results on  the asymptotic behavior.

\setcounter{equation}{0}
\section{Basic assumptions for linear viscoelastic fluids}
In this work we consider a viscoelastic fluid defined by means of the
constitutive equation for the Cauchy stress tensor%
\begin{equation*}
\mathbf{T}(\mathbf{x},t)=-p(\mathbf{x},t)\mathbf{I}+\mathbf{T}_{E}(\mathbf{E}_{r}^{t}(\mathbf{x})),
\end{equation*}%
where $p$ represents the \emph{pressure}, $\mathbf{I}$\ denotes the unit second
order tensor and the \emph{extra stress} $\mathbf{T}_{E}$\ is a function of the \emph{relative strain history}
$
\mathbf{E}_{r}^{t}(\mathbf{x},s)=\mathbf{E}(\mathbf{x},t-s)-\mathbf{E}(\mathbf{x},t)
$
at any fixed point of the material (\cite{Joseph1990}).
These fluids are described  by the classical Boltzmann-Volterra constitutive equation between the current value of the extra stress
 $\mathbf{T}_{E}(\mathbf{x},t)$
and the relative strain history $\mathbf{E}_r^{t}(\mathbf{x},s)$
\begin{equation}
\mathbf{T}_E(t)=2\int_{0}^{+\infty }\mu
^{\prime }( \mathbf{x},s) \mathbf{E}^{t}_r(\mathbf{x},s) \,ds.  \label{2.1}
\end{equation}%
Here  $\mu ^{\prime }$ is a constitutive function,
called \emph{memory kernel}, such that the \emph{shear relaxation function}%
\begin{equation}
\mu (\mathbf{x},s)=-\int_{s}^{+\infty }\mu ^{\prime }(\mathbf{x},\xi )d\xi \qquad  s\geq 0\notag
%\label{2.2}
\end{equation}%
belongs to $L^{1}
(\mathbb{R}^{+}; L^\infty(\Omega))$.
As proved in \cite{Fabrizio_Lazzari93}, the thermodynamic principles provide, almost everywhere in $\Omega$, the following restriction on its Fourier cosine transform:
\begin{equation}
\mu _{c}(\mathbf{x},\omega )>0\qquad \forall \omega \in\mathbb{R}^{+}.  \label{2.21}
\end{equation}
For these materials the physical state at time $t$ is identified through the  \emph{mass density} $\rho$  and  the relative strain history $\mathbf{E}_r^{t}(\mathbf{x},\cdot)$ at the time $t$.

In the following, we consider incompressible fluids, for which we have
$
\nabla \cdot \mathbf{v}=0$,
where $\mathbf{v}$\ denotes the \emph{velocity};  therefore, the state is defined
only by  means of the relative strain history.

By introducing  the vector space of the admissible relative strain
histories
\begin{equation}%\label{Gamma}
\Gamma _{r}=\left\{ {\bf E}_{r}^{t}: \Omega\times\mathbb{R}^{++}\rightarrow
Sym;\;\left| \int_{0}^{+\infty }\int_\Omega\mu ^{\prime }(\mathbf{x},\xi +\tau ){\bf
E}_{r}^{t}(\mathbf{x},\xi )d\mathbf{x}d\xi \right| <+\infty \,,\quad \forall \tau \geq
0\right\},\notag%
\end{equation}
it is possible to give the following equivalence relation (see \cite{fab5}).
\begin{defin}  \label{Def1}{\rm Let $\mathbf{x}\in\Omega$. Two relative strain histories ${\bf E}_{r_j}^t(\mathbf{x},\cdot)$,
$(j=1,2)$
are said to be equivalent if and only if
\begin{equation}% \label{equivalenza2}
\int_{0}^{+\infty }\mu ^{\prime }(\mathbf{x},\xi +\tau )\left[ {\bf E}
_{r_{1}}^{t}(\mathbf{x},\xi )-{\bf E}_{r_{2}}^{t}(\mathbf{x},\xi )\right] d\xi ={\bf 0}\,,
\qquad  \forall\tau >0 .\notag%
\end{equation}}
\end{defin}
Let us introduce
\begin{equation}  \label{I}
{\bf \breve{I}}^{t}(\mathbf{x},\tau)=-2\int_{0}^{+\infty }\mu ^{\prime }(\mathbf{x},\xi +\tau ){\bf E}
_{r}^{\mathbf{x},t}(\xi )d\xi \,.
\end{equation}
 As a consequence of  Definition \ref{Def1},  ${\bf \breve{I}}^{t}$  characterizes the equivalence class in the space of the admissible relative strain
histories and hence we will call it the \emph{minimal state}.

\setcounter{equation}{0}
\section{Application to the quasi-static  problem}
\setcounter{equation}{0}

In this section we study the quasi-static  problem for  incompressible %homogeneous
linear viscoelastic fluids using the minimal state $\mathbf{\breve{I}}^{t}$. In fact, thanks to (\ref{2.1}) and (\ref{I}), the extra stress can be written in the following manner
$$
\mathbf{T}_{E}(\mathbf{x},t)=-\mathbf{\breve{I}}^{t}(\mathbf{x},0)=2\int_{0}^{t }\mu (\mathbf{x},s)\mathbf{\dot{E}}^{t
}(\mathbf{x},s)ds-\mathbf{\breve{I}}^{0}(\mathbf{x},t)
$$
where $\mathbf{\breve{I}}^{0}$ is related to the initial past history and therefore is a known function.

Let $\Omega$ be a bounded domain in $\mathbb{R}^3$ with smooth boundary $\partial\Omega$. The linear approximation of the quasi-static boundary value problem with Dirichlet conditions is
\begin{equation}\label{8.1}
0={-\nabla p(\mathbf{x},t)+\nabla \cdot
\int_{0}^{t}\mu (\mathbf{x},s)\nabla \mathbf{v}^{t}(\mathbf{x},s)ds-\nabla
\cdot \mathbf{\breve{I}}^{0}(\mathbf{x},t)+   \mathbf{f}(\mathbf{x},t) }
\end{equation}
\begin{equation}
\nabla \cdot \mathbf{v}(\mathbf{x},t)=0,\quad \mathbf{v}(\mathbf{x},t)_{|\partial\Omega}=\mathbf{0}.%\quad [ \mathbf{v}(\mathbf{x},0)=\mathbf{0} \hbox {\bf\Large{ serve?}}
\label{8.11}
\end{equation}
In order to give a precise formulation of problem (\ref{8.1}) -- (\ref{8.11}) we introduce the space
$$J(\Omega)=\left\{ \mathbf{v}\in C_0^\infty(\Omega);\;\nabla \cdot \mathbf{v}=0\right\} $$
and denote by $\overset{\!\circ}{{L}^{2}}(\Omega )$ and
$\overset{\!\circ}{{H}_{0}^{1}}(\Omega )$ the closure of $J(\Omega)$ in the $L^2$ and $H^1$ norms respectively. Moreover we consider the spaces
\begin{equation}
\mathcal{H}_{\mu }(\mathbb{R}^{+},\Omega )=\left\{ \mathbf{v} \in  L_{loc}^{2}(\mathbb{R}
^{+};\overset{\!\circ}{{H}_{0}^{1}}(\Omega ));
\int_{0}^{+\infty }\!\!\!\!\int_{0}^{+\infty }\!\!\!\!\int_{\Omega }\mu (\mathbf{x}%
,\mid \tau -\tau ^{\prime }\mid )\nabla \mathbf{v}(\mathbf{x},\tau )\cdot
\nabla \mathbf{v}(\mathbf{x},\tau ^{\prime })d\mathbf{x}d\tau d\tau ^{\prime
} <+\infty\right\},
\end{equation}
\begin{equation}
\mathcal{S}_{\mu }(\mathbb{R}^{+},\Omega ) =\left\{ \mathbf{\breve{I}}^{0}\in L_{loc}^{2}(\mathbb{R}
^{+};{{L}^{2}}(\Omega ));  \int_{0}^{+\infty }\!\!\!\!\int_{0 }^{+\infty }\!\!\!\!\int_{\Omega }\tilde{%
\mu}(\mathbf{x},\mid \tau -\tau^{\prime}\mid)\mathbf{\breve{I}}^{0}(%
\mathbf{x},\tau )\cdot \mathbf{\breve{I}}^{0}(\mathbf{x},\tau^{\prime })d%
\mathbf{x}d\tau d\tau ^{\prime }
 <+\infty \right\},
\end{equation}
where $\tilde{\mu}$ is defined by
$$
\frac14\int_{-\infty }^{+\infty }\mu (\mathbf{x},\mid \tau -\tau ^{\prime }\mid )%
\tilde{\mu}(\mathbf{x},\mid \tau ^{\prime }\mid )d\tau ^{\prime }=\delta
(\tau ).$$
\begin{rem}
\rm{If the kernel $\mu \in L^\infty(\Omega,L^{1}(\mathbb{R}^{+}))$ satisfies the thermodynamic condition (\ref{2.21}) almost everywhere in $\Omega$, then $\mathcal{H}_{\mu }$ and $\mathcal{{S}}_{\mu }$ are Hilbert spaces.
In fact it is possible to define the spaces $\mathcal{H}_{\mu }$ and $\mathcal{{S}}_{\mu }$ in the frequency  domain (see \cite{fab5})
by observing that
\begin{equation}\label{H}
 \int_{0}^{+\infty }\!\!\!\!\int_{0}^{+\infty }\!\!\!\!\int_{\Omega }\mu (\mathbf{x}%
,\mid \tau -\tau ^{\prime }\mid )\nabla \mathbf{v}(\mathbf{x},\tau )\cdot
\nabla \mathbf{v}(\mathbf{x},\tau ^{\prime })d\mathbf{x}d\tau d\tau ^{\prime
}=\frac{1}{\pi }\int_{-\infty }^{+\infty }\!\!\!\!\int_{\Omega }\mu _{c}(
\mathbf{x},\omega )|\nabla \mathbf{v}_{F}(\mathbf{x},\omega )|^2 d\mathbf{x}d\omega \,,
\end{equation}
\begin{equation}\label{S}
\int_{0}^{+\infty }\!\!\!\!\int_{0}^{+\infty }\!\!\!\!\int_{\Omega }\tilde{
\mu}(\mathbf{x},\mid \tau -\tau ^{\prime }\mid )\mathbf{\breve{I}}^{0}(
\mathbf{x},\tau )\cdot \mathbf{\breve{I}}^{0}(\mathbf{x},\tau ^{\prime })d
\mathbf{x}d\tau d\tau ^{\prime }
=\frac{1}{\pi }\int_{-\infty }^{+\infty }\!\!\!\!\int_{\Omega }\frac1{\mu _{c}(\mathbf{x},\omega )}%
|\mathbf{\breve{I}}{}_{F}^{0}(\mathbf{x},\omega )|^2d\mathbf{x}d\omega\, ,
\end{equation}
where the index $_F$ denotes the Fourier transform.

We finally recall that the left-hand side of (\ref{H}) is the expression of the minimal free energy introduced in \cite{Breuer1} and  \cite{Breuer2}.}
\end{rem}
\begin{defin}
\label{D8.1}\rm{\ A function $\mathbf{v}\in \mathcal{H}_{\mu }(\mathbb{R}^{+},\Omega )$ is said to be a weak solution in the sense of the Virtual Power Principle of the problem (\ref{8.1}) -- (\ref{8.11}) %with $\mathbf{\breve{I}}^{0}\in \mathcal{S}_{\mu }(\mathbb{R}^{+},\Omega )$,
%and vanishing external force $\mathbf {f}$,
 if %it satisfies the equation
\begin{eqnarray}
2\int_{0}^{+\infty }\int_{\Omega }\int_{0}^{t}\mu (\mathbf{x},t-s)\nabla \mathbf{v%
}(\mathbf{x},s)ds\cdot \nabla \mathbf{w}(\mathbf{x},t) d\mathbf{x}%
dt\quad &&  \notag \\
=\int_{0}^{+\infty }\int_{\Omega }\mathbf{%
\breve{I}}^{0}(\mathbf{x},t)\cdot \nabla \mathbf{w}(\mathbf{x},t) d\mathbf{x}
dt  -\int_{0}^{+\infty }\int_{\Omega }\mathbf{f}(\mathbf{x},t)\cdot \mathbf{w}(\mathbf{x},t) d\mathbf{x} dt &&  \label{8.10}
\end{eqnarray}%
for any $\mathbf{w}\in \mathcal{H}_{\mu }(\mathbb{R}^{+},\Omega )$}.
% and $\underset{t\rightarrow +\infty }{\lim } \mathbf{w}(\mathbf{x},t)=\mathbf{0}$.}
\end{defin}

\begin{rem}
\rm{ Given a vector ${\mathbf{f}}$, let ${{\nabla\times\mathbf{a}}}$ and $\nabla \phi$ be
its solenoidal and irrotational components, respectively, in
the Helmholtz decomposition, i.e.  ${\mathbf{f}} ={\nabla\times\mathbf{a}}  +\nabla \phi$. Since
$\mathbf{w}\in \mathcal{H}_{\mu }(\mathbb{R}^{+},\Omega )$, the last integral in (\ref{8.10})
can be rewritten as
$$\int_{0}^{+\infty }\int_{\Omega } \mathbf{f}(\mathbf{x},t)\cdot \mathbf{w}(\mathbf{x},t)d\mathbf{x} dt  =\int_{0}^{+\infty }\int_{\Omega }\nabla\times \mathbf{a}(\mathbf{x},t)\cdot \mathbf{w}(\mathbf{x},t)d\mathbf{x} dt.
$$
Moreover, introducing the skew tensor  $\mathbf{A}$, defined through the relation
$\mathbf{A}\mathbf{w}=\mathbf{a}\times\mathbf{w}$, we have $\nabla\cdot \mathbf{A}= \nabla\times \mathbf{a}$ so that
$$\int_{0}^{+\infty }\int_{\Omega } \mathbf{f}(\mathbf{x},t)\cdot \mathbf{w}(\mathbf{x},t)d\mathbf{x} dt  =-\int_{0}^{+\infty }\int_{\Omega } \mathbf{A}(\mathbf{x},t)\cdot\nabla\mathbf{w}(\mathbf{x},t)d\mathbf{x} dt.
$$
}
\end{rem}
By virtue of the previous remark, relation (\ref{8.10}) becomes
\begin{equation}
\int_{0}^{+\infty }\int_{\Omega }\int_{0}^{t}\mu (\mathbf{x},t-s)\nabla \mathbf{v%
}(\mathbf{x},s)ds\cdot \nabla \mathbf{w}(\mathbf{x},t) d\mathbf{x}%
dt
=\int_{0}^{+\infty }\int_{\Omega }\mathbf{%
\breve{J}}^{0}(\mathbf{x},t)\cdot \nabla \mathbf{w}(\mathbf{x},t) d\mathbf{x}
dt ,   \label{8.10a}
\end{equation}
where $\mathbf{\breve{J}}^{0}=\mathbf{\breve{I}}^{0}-\mathbf{A}$.
\begin{teo} \label{T8.1}
Problem
{\rm{(\ref{8.1}) -- (\ref{8.11})}} with $\mathbf{\breve{J}}^{0}\in {\mathcal{S}}_{\mu }(\mathbb{R}^{+},\Omega )$ admits a unique solution,
%$\mathbf{v}\in \mathcal{H}_\mu(\mathbb{R}^{+},\Omega )$
according to Definition {\rm \ref {D8.1}}, if $\mu \in L^{1}(\mathbb{R}
^{+},L^\infty(\Omega) )$ and  {\rm{(\ref{2.21})}}  holds almost everywhere in $\Omega$.
 \end{teo}
%the kernel
%$$\mu \in L^{1}(\mathbb{R}^{+},L^\infty(\Omega) )\,,\;  \mu^\prime \in L^{1}(\mathbb{R}^{+},L^\infty(\Omega) )\cap L^{2}(\mathbb{R}^{+},L^\infty(\Omega) )\,,\; \mu^{\prime\prime} \in L^{1}(\mathbb{R}^{+},L^\infty(\Omega) )$$and conditions {\rm{(\ref{2.21})}} and  {\rm{(\ref{mu'0})}}
\textit{Proof. }
Let us consider the Fourier transform of system (\ref{8.1}) -- (\ref{8.11})
\begin{equation}\label{8.12}
\begin{split}
&\nabla \cdot \lbrack \mu _{F}(%
\mathbf{x},\omega )\nabla \mathbf{v}_{F}(\mathbf{x},\omega )]-
\nabla \left[p_{F}(\mathbf{x},\omega )-\phi_{F}(\mathbf{x},\omega )\right]
=\nabla \cdot \mathbf{\breve{J}}_{F}^{0}(\mathbf{x},\omega)  \\
&\qquad\nabla \cdot \mathbf{v}_{F}(\mathbf{x},\omega )=0,   \quad \mathbf{v}_{F}(\mathbf{x},\omega )\mid _{\partial\Omega }=\mathbf{0}
\end{split}
\end{equation}
and introduce the bilinear form
\begin{equation}
b_\omega(\mathbf{v}_{F},\mathbf{w}_{F})
=\int_{\Omega }\mu _{c}(\mathbf{x},\omega )\nabla \mathbf{v}_{F}(\mathbf{x}, \omega)\left[ \nabla \mathbf{w}_{F}(\mathbf{x},\omega )\right] ^{\ast }d%
\mathbf{x}  \notag%\label{8.15}
\end{equation}
for any fixed $\omega \in \mathbb{R}$, where the index $^*$ denotes the complex conjugate.
The hypotheses on the kernel $\mu$ ensure that $b_\omega$  is bounded  and coercive in $\overset{\!\circ}{{H}_{0}^{1}}(\Omega )$ since
\begin{equation}
k_1(\omega )\|\nabla \mathbf{v}_{F}(\omega)\|^2 \leq  b_\omega(\mathbf{v}_{F},\mathbf{v}_{F})\leq
k_2(\omega )\|\nabla \mathbf{v}_{F}(\omega)\|^2,
 \notag%\label{8.16}
\end{equation} where $k_1(\omega )$ and  $k_2(\omega )$ are  the essential  infimum and the essential supremum  of $\mu _{c}(\mathbf{x},\omega )$ on $%
\Omega $ respectively.

Therefore, thanks to  the Lax-Milgram theorem, system (\ref{8.12}) admits, for any fixed $\omega \in\R$,  one and only one solution belonging to $ \overset{\!\circ}{{H}_{0}^{1}}(\Omega )$ if the supply $\nabla \cdot \mathbf{\breve{J}}_{F}^{0}(\cdot,\omega)$
%$\nabla \cdot \mathbf{\breve{J}}_{+}^{0}(\cdot,\omega)\in \overset{\!\!\!\!\!\circ}{{H}^{-1}}(\Omega )$,
belongs to the dual space of  $\overset{\!\circ}{{H}_{0}^{1}}(\Omega )$.

Moreover, if we rewrite  (\ref{8.10a}), by virtue  Plancherel's theorem, as follows
\begin{equation}
\int_{-\infty }^{+\infty }\int_{\Omega }\mu_{F}(\mathbf{x},\omega )\nabla \mathbf{v}_{F}(\mathbf{x},\omega )\cdot \left[\nabla \mathbf{w}_{F}(\mathbf{x},\omega )\right] ^{\ast} d\mathbf{x}d\omega
 =\int_{-\infty }^{+\infty }\int_{\Omega }
\mathbf{\breve{J}}_{F}^{0}(\mathbf{x},\omega )\cdot \left[ \nabla \mathbf{w}_{F}(\mathbf{x},\omega )\right] ^{\ast }d\mathbf{x}d\omega  \label{8.18}
\end{equation}
and choose  $\mathbf{w}_F=\mathbf{v}_F$ in  (\ref{8.18}), we obtain
\begin{eqnarray}
&&\int_{-\infty }^{+\infty }\int_{\Omega }\mu _{c}(\mathbf{x},\omega )\mid
\nabla \mathbf{v}_{F}(\mathbf{x},\omega )\mid ^{2}d\mathbf{x}d\omega
 =\int_{-\infty }^{+\infty }\int_{\Omega }\frac{\mathbf{\breve{J}}_{F}^{0}(\mathbf{x},\omega )}{\sqrt{\mu _{c}(\mathbf{x}%
,\omega )}}
\cdot \sqrt{\mu _{c}(\mathbf{x}%
,\omega )}\left[ \nabla \mathbf{v}_{F}(\mathbf{x},\omega )\right]
^{\ast }d\mathbf{x}d\omega  \notag \\
&&\quad \leq \left[ \int_{-\infty }^{+\infty }\int_{\Omega }\frac1{\mu _{c}(
\mathbf{x},\omega )}\mid \mathbf{\breve{J}}_{F}^{0}(\mathbf{x},\omega )\mid
^{2}d\mathbf{x}d\omega \right] ^{1/2}
 \left[ \int_{-\infty }^{+\infty }\int_{\Omega
}\mu _{c}(\mathbf{x},\omega )\mid \nabla \mathbf{v}_{F}(\mathbf{x},\omega
)\mid ^{2}d\mathbf{x}d\omega \right] ^{1/2}.   \notag%\label{8.19}
\end{eqnarray}%
Hence, thanks to (\ref{H}) and (\ref{S}), $\mathbf{v}\in \mathcal{H}_{\mu }(\mathbb{R}^{+},\Omega )$ if $\mathbf{\breve{J}}^{0}\in \mathcal{S}_{\mu }(\mathbb{R}^{+},\Omega )$
because
\begin{equation}\int_{-\infty }^{+\infty }\int_{\Omega }\mu _{c}(\mathbf{x},\omega )\mid
\nabla \mathbf{v}_{F}(\mathbf{x},\omega )\mid ^{2}d\mathbf{x}d\omega
\leq
\int_{-\infty }^{+\infty }\int_{\Omega }\frac1{\mu _{c}(\mathbf{x},\omega
)}
\mid\mathbf{\breve{J}}_{F}^{0}(\mathbf{x},\omega )\mid ^{2}d\mathbf{x}%
d\omega .  \label{8.20}
\end{equation}
Since  $\mathbf{\breve{J}}^{0}=\mathbf{\breve{I}}^{0}-\mathbf{A}$,
where $\nabla\cdot\mathbf{A} $ is the solenoidal part of the external force $\mathbf{f}$, we conclude that
 the virtual power solution of  problem   (\ref{8.1}) -- (\ref{8.11})
belongs to $\mathcal{H}_{\mu}(\mathbb{R}^{+};\Omega )$ if the initial datum $\mathbf{\breve{I}}^{0}\in\mathcal{S}_{\mu}(\mathbb{R}^{+};\Omega )$
and the solenoidal component  $\mathbf{f}_s$ of $\mathbf{f}$
belongs to
the space $L_{loc}^{2}(\mathbb{R}^{+};H^{\prime}(\Omega ))$ where $H^{\prime}(\Omega )$ is the dual of $\overset{\!\circ}{{H}_{0}^{1}}(\Omega )$ and there exists a skew tensor
$\mathbf{A}\in \mathcal{S}_{\mu }(\mathbb{R}^{+},\Omega ) $ such that  $\mathbf{f}_s= \nabla\cdot \mathbf{A}$.
\hfill $\Box $

\setcounter{equation}{0}
\section{Asymptotic behavior}
Equation (\ref{8.1}), in absence of external forces,
% $$\nabla\cdot\mathbf{T}_E(\mathbf{x},t)=\nabla p(\mathbf{x},t)$$
can be rewritten in terms of  the minimal state (\ref{I}) as follows
%\mathbf{\breve{I}}^{t}(\mathbf{x},\cdot)
 \begin{equation}\label{exp1}
 \nabla\cdot\mathbf{\breve{I}}^{t}(\mathbf{x},0)+\nabla p(\mathbf{x},t)=0.
\end{equation}
 It is therefore necessary to assign the law governing the evolution in time of $\mathbf{\breve{I}}^{t}$ which, taking into account definition (\ref{I}), is given by
   \begin{equation}\label{Ipunto}
 \frac{\partial}{\partial t} \mathbf{\breve{I}}^{t}(\mathbf{x},\tau)  =   \frac{\partial}{\partial \tau}\mathbf{\breve{I}}^{t}(\mathbf{x},\tau)  - 2 \mu(\mathbf{x},\tau) \nabla \mathbf{v} (\mathbf{x},t)\,,\qquad\mathbf {\breve{I}}^{0}(\mathbf{x},\tau) =\mathbf {{I}}_{0}(\mathbf{x},\tau),
\end{equation}
where $\mathbf {{I}}_{0}(\mathbf{x},\tau)$ is a known function on $\Omega\times \mathbb{R}^{+}$.

Theorem \ref{T8.1} ensures that problem  (\ref{exp1}),  (\ref{Ipunto}) and (\ref{8.11}) admits  a unique solution, according   to Definition 4.1, whenever   the initial datum  $\mathbf {{I}}_{0}$ belongs
to $\mathcal{S}_\mu(\mathbb{R}^{+}, \Omega)$.
In this section we will study the connection between the asymptotic behavior  of this solution and that of the  memory kernel $\mu$.
To this aim we restrict ourselves to  memory kernels satisfying almost everywhere in $\Omega$ the following restrictions
\begin{equation}\label{diss4}
\mu^{\prime} (\mathbf{x},t)<0\,, \quad\mu^{\prime\prime}(\mathbf{x},t)\geq 0 \,,\quad
\mu^{\prime\prime} (\mathbf{x},t)) \geq -\xi(t)\mu^{\prime} (\mathbf{x},t))\,, \qquad t\geq0
\end{equation}
where $\xi$ is a positive, non-increasing differential function.
%such that it exists a positive constant $\kappa$ for which \begin{equation}\label{diss5}  -\xi^{\prime}(t)   \leq \kappa \xi (t)\,,\qquad t\geq0.\end{equation}

Examples of such kernels can be found for example in \cite{Messaoudi2008}; in particular,
a kernel presents an exponential or polynomial decay  when  $\xi$ is a constant function or
$\xi(t)= c(1+t)^{-1}$, respectively.

Let us now consider problem  (\ref{exp1}),  (\ref{Ipunto}) and (\ref{8.11})  with initial datum
$\mathbf {{I}}_{0}$ belonging to the subspace%
\begin {footnote} {We recall that, as proved in \cite{Fabrizio_Golden2002}$,  \mathcal{H}_\mu(\mathbb{R}^{+}, \Omega)$ is the space where the minimal free energy is defined, while
$\mathcal{F}_\mu(\mathbb{R}^{+}, \Omega)$  is the domain of the free energy introduced in \cite{fab0}.  Therefore $ \mathcal{F}_\mu(\mathbb{R}^{+}, \Omega)\subset  \mathcal{H}_\mu(\mathbb{R}^{+}, \Omega)$.}
\end{footnote}
of $ \mathcal{S}_\mu(\mathbb{R}^{+}, \Omega)$ defined by
\begin{equation}
\mathcal{F}_{\mu }(\mathbb{R}^{+},\Omega ) =\left\{ \mathbf{\breve{I}}^{0}\in \mathcal{S}_{\mu }(\mathbb{R}^{+},\Omega ) ;
 \int_0^\infty \int_\Omega\frac{1}{-\mu^{\prime} (\mathbf{x},\tau)} \left|  \frac{\partial}{\partial \tau}  \mathbf {\breve{I}}^{0}(\mathbf{x},\tau)\right|^2  d\mathbf{x} d\tau<\infty
 \right\}
\end{equation}
and introduce the energy functional
\begin{equation}\label{Psi}
\Psi(t)= \Psi(   \mathbf {\breve{I}}^{t}) =-\frac14\int_0^\infty \int_\Omega\frac{1}{\mu^{\prime} (\mathbf{x},\tau)} \left|  \frac{\partial}{\partial \tau}  \mathbf {\breve{I}}^{t}(\mathbf{x},\tau)\right|^2  d\mathbf{x} d\tau.
\end{equation}
If   $     \mathbf {\breve{I}}^{t}    \in \mathcal{F}_{\mu }(\mathbb{R}^{+},\Omega ) 	$ this   functional  satisfies
\begin{equation}\label{Psiderivata}
\frac{d}{d t} \Psi(t)=-\int_\Omega   \mathbf {\breve{I}}^{t}(\mathbf{x},0)\cdot \nabla   \mathbf {v}(\mathbf{x},t)d\mathbf{x} -\frac14\int_0^\infty \!\!\!\int_\Omega\frac{\mu^{\prime\prime}(\mathbf{x},\tau)}{\left[\mu^{\prime} (\mathbf{x},\tau)\right]^2} \left|  \frac{\partial}{\partial \tau}  \mathbf {\breve{I}}^{t}(\mathbf{x},\tau)\right|^2  d\mathbf{x} d\tau
+
\frac14\int_\Omega\frac{1}{\mu^{\prime} (\mathbf{x},0)} \left|  \frac{\partial}{\partial \tau}  \mathbf {\breve{I}}^{t}(\mathbf{x},0)\right|^2  d\mathbf{x}\,;\qquad\qquad\qquad\qquad\qquad
\end{equation}
moreover, fixed $T_0>0$, there exists $\alpha_{T_0}>1$ such that
\begin{equation}\label{maggiorazione}
 \Psi(t)\leq-\frac{ \alpha_{T_0}}4\int_0^t \int_\Omega\frac{1}{\mu^{\prime} (\mathbf{x},\tau)} \left|  \frac{\partial}{\partial \tau}  \mathbf {\breve{I}}^{t}(\mathbf{x},\tau)\right|^2  d\mathbf{x} d\tau\qquad\forall t>T_0.
\end{equation}
Let $ \mathbf {\breve{I}}^{t}$ be a solution of  (\ref{exp1}),  (\ref{Ipunto}) and (\ref{8.11})  with initial datum
$\mathbf {{I}}_{0}\in \mathcal{F}_{\mu }(\mathbb{R}^{+},\Omega )$. As a consequence of
 (\ref{exp1}),  (\ref{8.11}), (\ref{diss4}) and (\ref{Psiderivata})
we obtain
\begin{equation}\label{Psiderivata1}
\frac{d}{d t} \Psi(t)\leq \frac14\int_0^\infty \int_\Omega\frac{\xi(\tau)}{\mu^{\prime} (\mathbf{x},\tau)} \left|  \frac{\partial}{\partial \tau}  \mathbf {\breve{I}}^{t}(\mathbf{x},\tau)\right|^2  d\mathbf{x} d\tau
\leq \frac14\int_0^t \int_\Omega\frac{\xi(\tau)}{\mu^{\prime} (\mathbf{x},\tau)} \left|  \frac{\partial}{\partial \tau}  \mathbf {\breve{I}}^{t}(\mathbf{x},\tau)\right|^2  d\mathbf{x} d\tau\leq0.\end{equation}
Finally, if  $ t>T_0$,  the properties of $\xi$ and the inequality (\ref{maggiorazione}) yield
\begin{equation}\label{Psiderivata2}
\frac{d}{d t} \Psi(t)\leq \frac{\xi(t)}4\int_0^t \int_\Omega\frac{1}{\mu^{\prime} (\mathbf{x},\tau)} \left|  \frac{\partial}{\partial \tau}  \mathbf {\breve{I}}^{t}(\mathbf{x},\tau)\right|^2  d\mathbf{x} d\tau
\leq  \frac{\xi(t)}{4\alpha_{T_0}}\int_0^\infty \!\!\!\int_\Omega\frac{1}{\mu^{\prime} (\mathbf{x},\tau)} \left|  \frac{\partial}{\partial \tau}  \mathbf {\breve{I}}^{t}(\mathbf{x},\tau)\right|^2  d\mathbf{x} d\tau= - \frac{\xi(t)}{\alpha_{T_0}}  \Psi(t)
\end{equation}
and the integration of (\ref{Psiderivata2}) gives
\begin{equation}\label{stima}
 \Psi(t)\leq   \Psi(T_0)  \exp\left[- \frac{1}{\alpha_{T_0}}\int_{T_0}^t\xi(s)ds\right]
 \leq   \Psi(\mathbf {{I}}_{0})  \exp\left[- \frac{1}{\alpha_{T_0}}\int_{T_0}^t\xi(s)ds\right]
 \,,\quad t>T_0.
 \end{equation}
 We conclude this section by stating the following theorem
 \begin {teo}\label{teo2}
Let $ \mathbf{v} $ be a virtual work solution of problem $(\ref{8.1})-(\ref{8.11})$ with a vanishing external source and $ \mathbf{\breve{I}}^{0}\in \mathcal{F}_{\mu }(\mathbb{R}^{+},\Omega )$. If $\mu$ satisfies  $(\ref {diss4})$,
then, for $T_0>0$, there exist two positive constants $\alpha_{T_0}$ and $\beta_{T_0}$ such that %satisfies
\begin{equation}\label{stima1}
 \Psi(t)\leq  \beta_{T_0}  \Psi({0})  \exp\left[- \frac{1}{\alpha_{T_0}} \int_{0}^t\xi(s)ds\right]
 \,,\quad t>T_0.
 \end{equation}
 \end{teo}
\begin {cor}
Under the hypotheses of {\rm Theorem \ref{teo2}}, the energy functional $(\ref{Psi})$ exponentially (polinomially) decays if the memory kernel $\mu$ exponentially (polinomially) decays in time. \end{cor}
\textit{Proof. } It is easy to show that if $\xi$ is constant in time, then (\ref{diss4}) assures the exponential decay of $\mu$, while  (\ref{stima1}) yields the exponential decay of the energy.

\noindent On the other hand, if $\xi(t)=c(1+t)^{-1}$, then $-\mu^\prime(t)=O((1+t)^{-c})$ and from (\ref{stima1}) we obtain
$$ \Psi(t)\leq  \beta_{T_0}  \Psi({0}) (1+t)^{-c / \alpha_{T_0}}
 \,,\quad t>T_0.
$$\hfill$\Box$
\bibliographystyle{elsarticle-harv}
\bibliography{biblio}

\end{document}